%% file: arxiv.tex
\documentclass[1opt,twocolumn,letter]{IEEEtran}
\usepackage{graphicx}
\DeclareGraphicsExtensions{.eps}
\graphicspath{{figures/}}
\usepackage[tight]{subfigure}
\usepackage{fancyhdr}
\usepackage{fancyvrb}
\usepackage{boxedminipage}
\usepackage{framed}

\usepackage{times}

\usepackage[ruled,slide, vlined]{algorithm2e}
\usepackage{multicol}

\newcommand{\ie}{\textit{i.e.},\xspace}
\newcommand{\eg}{\textit{e.g.},\xspace}
\newcommand{\ignore}[1]{{}}

\makeatletter \newcommand \listoftodos{\section*{Todo list} \@starttoc{tdo}}
\newcommand\l@todo[2]{
  \par\noindent \textit{#2}, \parbox{10cm}{#1}\par
} 
\makeatother

\begin{document}

\title{OS-level Failure Injection with SystemTap}

\author{
\begin{tabular}{cc}
Camille Coti\dag & Nicolas Greneche\dag\ddag\\
\end{tabular}\\
\dag {\it LIPN, CNRS, UMR 7030} \\
\ddag {\it DSI}\\
\textit{Universit\'e Paris 13, Sorbonne Paris Cit\'e}\\
\textit{F-93430, Villetaneuse, France}\\
\textit{\{first.last\}@univ-paris13.fr}
}

\date{} 

\maketitle
\begin{abstract}
Failure injection in distributed systems has been an important issue
to experiment with robust, resilient distributed systems. In order to
reproduce real-life conditions, parts of the application must be
killed without letting the operating system close the existing network
communications in a "clean" way. When a process is simply killed, the
OS closes them. SystemTap is a an infrastructure that probes the Linux
kernel's internal calls. If processes are killed at kernel-level, they
can be destroyed without letting the OS do anything else. In this
paper, we present a kernel-level failure injection system based on
SystemTap. We present how it can be used to implement deterministic
and probabilistic failure scenarios. 
\end{abstract}


\section{Introduction}
\label{sec:intro}

Failure injection in distributed systems has been an important issue
to experiment with robust, resilient distributed systems. People who develop
fault-tolerant applications must be able to test their software in faulty
conditions, \ie, with realistic failures. For instance, the fault-tolerant
implementation of the MPI standard MPICH-V \cite{SC02} aims at providing an
execution environment that can survive failures. It was tested using a
failure-injection system and a specific fault scenario exhibited a rare bug in a
precise situation of two consecutive failures \cite{HoarauLHRTC06}. This tool
helped the developers of MPICH-V finding out this bug and, more important,
fixing it. 

In a similar way, failure detectors need to be strained and tested in real-life
situations \cite{ACT}. Traditional failure injection tools kill
processes. However, when a process is killed, the operating system of the
machine closes all the network and system sockets in a ``clean'' way. For
instance, all the TCP connections are closed according to the TCP
protocol. However, this is not a realistic situation. In real life, most
failures follow the \emph{fail-stop semantics}: when a component fails, it
simply stops doing anything \cite{Powell92, ScSc83}. The crashed component can
be a process or a communication channel. The cause of this failure can be an
electric failure, a cable cut, a scheduling bug... In any case, the failed
component behaves normally and then halts. As a matter of fact, no warning is
issued before the crash happens. As a consequence, the operating system cannot
close the sockets nor any such thing. 

In this paper, we present a method to inject failures in applications \emph{at
  kernel-level}; therefore, the operating system cannot interfere with the
application and the sockets are not closed. Our approach is based on two Linux
tools: \emph{SystemTap} and \emph{control groups} (cgroups). We also explain how
this approach can be extended to other crash semantics, such as replication and
omission of network packets.

The remainder of this paper is organized as follows. In section
\ref{sec:related}, we present and analyze existing failure injection systems,
and we compare them with our approach. In section \ref{sec:fist}, we present our
approach for Failure Injection with SystemTap (FIST), and how it can be used to
inject failures in distributed applications. In section \ref{sec:scenarios}, we
explain how failure scenarios are implemented in practice with
FIST. Last, we conclude the presentation of this approach and we
mention perspectives for other kinds of failures in section 
\ref{sec:conclu}. 

\input{related}
\input{injection}

\section{Conclusion and perspective}
\label{sec:conclu}

In this paper, we have presented FIST, a failure-injection technique
that takes advantage of recent Linux kernel features, namely, cgroups
and SystemTap. This technique injects highly realistic failures,
taking advantage of the fact that it kills processes directly in the
kernel's task scheduler, making them die without notice. We have
presented a method to use it to implement deterministic and
probabilistic failure scenarios. 

\subsection{Limitations and how they can be overcome}

The major limitation of this approach is that the SystemTap module
must, of course, be executed with super-user privileges. This can be
overcome by two approaches. The most generic one is to use {\tt
  sudo}. The administrator of the machine can allow it on the {\tt
  modprobe} command only. As a consequence, the only thing that users
would be able to do is to insert their modules on the kernel. 

The other way is to work on an experimental testbed that provides a 
deployment software such as Kadeploy \cite{kadeploy}, such as the
Grid'5000 platform \cite{grid5000}. Kadeploy allows users to deploy
their own system image on the nodes they have reserved and therefore
have (temporarily) their own root access on these nodes. 

\subsection{Using this method to implement other failure models}

In this work we have focused in fail-stop failures. However, it can be
easily extended to other semantics. For instance, probes can be added
in the network stack. Messages can be dropped to inject
\emph{omission} failures, or sent twice to inject \emph{duplication}. 

\footnotesize
\bibliographystyle{plain}
\bibliography{fist}

\end{document}

%% file: related.tex
\section{Related works}
\label{sec:related}

Faults injection in parallel computing is more trick than in common
distributed systems. Developers can use a virtual HPC infrastructure
to check fault resilience of parallel code running on top of
parallelisation libraries (OpenMPI, OpenMP etc.)
\cite{HeraultLPQCJ09}. The libraries used for parallel computing are
closely related to hardware. For example, when Open MPI's run-time
daemon {\tt orted} is started, it selects a network medium (Ethernet,
InfiniBand, etc.) to perform communication between MPI
processes. Network media like Infiniband cannot be virtualized. 

In \cite{HoarauT05}, a daemon process is used on each node to apply a
fault injection policy in the instrumented processes. This approach
introduces additional processes on the host. Those processes interfere
with kernel scheduler in two ways: 
\begin{enumerate}
\item it consumes resources, such as CPU cycles and scheduling quanta;
\item more processes need to be handled by the scheduling policy.
\end{enumerate}
The scheduler behavior should not be altered by the fault injection
infrastructure in parallel computing libraries. For example,
parallel computing libraries like OpenMP \cite{chapman2008using}
interact with scheduler to optimize process affinity with CPU caches
etc \cite{hwloc}. Moreover, if a fault injection daemon fails, the
policy becomes inapplicable. 


%% file: injection.tex
\section{Failure injection}
\label{sec:fist}

\label{sec:fist:cgroups}

In common-Linux based systems, a running process is a kernel structure 
{\tt task\_struct}. This structure defines an integer variable {\tt
  pid} and a pointer to another {\tt task\_struct}. The {\tt pid} is
used by the kernel to identify the process in a non-ambiguous way. The
pointer refers the parent {\tt task\_struct}. To put it another way,
it refers to the father process that performed the fork/exec sequence
to spawn the child process. As a consequence, Linux process
organization can be seen as a tree with the {\tt initd} process as a
root.

\emph{Control groups} (Cgroups) are a major evolution of this
model. Without cgroups, resource usage policy can only be defined for
a targeted process and its potential childs. Cgroups allow to add a
process dynamicalle to one or more groups at each creation of a {\tt
  task\_struct}. Such a group is called a \emph{cgroup}. Resource
usage policy can target a cgroup instead of a single process. Cgroups
make it possible to apply the same resource usage policy on a group of
processes without wondering about the process tree structure. 
 
Cgroups are always used with a new improvement of Linux kernel:
\emph{name spaces}. Name spaces make it possible to have different
instances of some kernel objects. This provides powerful isolation
capabilities to Linux kernels. For instances, processes that belong to
a given cgroup can use their own instance of the IP stack. These two
components allow to run a lightweight sandbox under a strict resource
limitation policy \cite{merkel2014docker}.

\subsection{SystemTap}
\label{sec:fist:stap}

\emph{SystemTap} is a production-ready kernel auditing tool. An audit
policy is written in a high-level language. The main part of the
policy consist in definitions of \emph{probes}. A probe is an 
instrumentation point of the kernel (for example, a kernel
function). Systemtap relies on two kinds of probes: \emph{Kprobes} and
\emph{Jprobes}. Kprobes can trap at almost any kernel code address and
define a handler to execute code when the address is reached. A Kprobe
can be attached to any instruction in the kernel. Jprobes are inserted
at a kernel function's entry, providing a convenient access to its
arguments.  

Systemtap provides a compiler that transforms an auditing policy to a
loadable kernel module. When loaded, this kernel module registers all
the defined probes. Every time a probed function is reached, the defined
handler code is executed. Handler codes can use SystemTap's native
collection of functions or a guru mode. SystemTap's native functions are
high-level primitives implemented in \emph{tapsets} (quite similar to
libraries). The guru mode enables to pass rough C kernel code to handler
code. 

\subsection{Process supervision and control}
\label{sec:fist:supervision}

\emph{Process supervision} consists in collecting states and metrics
about a targeted process. \emph{Control} consist in performing actions
on a supervised process. Prior to the introduction of cgroups in the
Linux kernel, common ``master'' processes were used to supervise and
control child processes. The two main drawbacks of this model are the
following ones: 
\begin{enumerate}
\item if the master process fails, every supervised child process also
  fails. To mitigate this issue, the master's source code should be
  very short; 
\item a process can belong to only one supervision domain, since it
  has only one father.
\end{enumerate}

\subsection{Running distributed applications on FIST}
\label{sec:fist:run}

The idea behind FIST can be summarized as follows:
\begin{itemize}
\item a specific \emph{control group} is defined for processes that belong to
  the experiment; these are the processes that can be affected by intentional
  failures.
\item a \emph{SystemTap} script inserts a hook in the kernel's scheduler. When
  the scheduler is invoked to run a process, it checks whether this process
  belongs to the experimental cgroup. If it does, the failure injection
  scenario is followed to decide whether the process must be run normally or
  killed.
\end{itemize}

In practice, a process can be executed in a given cgroup by using the command
{\tt cgexec}. For instance, the command {\tt sleep} can be executed in the
cgroup {\tt xp} on all the mounted controllers using the following command:

\begin{Verbatim}[frame=single,fontsize=\relsize{-1}]
cgexec -g *:xp sleep 1
\end{Verbatim}

Distributed applications are often made of two parts: the run-time environment,
and the application processes themselves. In the case of MPI programs, the
application is supported by a distributed run-time environment which is made of
a set of processes (one on each machine used by the computation) that are
spawning, supporting and monitoring the application processes on their machine
\cite{ORTE, BGL00}. The failure detector is usually implemented in this run-time
environment. As a consequence, the processes of the run-time environment are the
ones that need to be executed in the experimental cgroup. For instance, Open MPI 
\cite{openmpi} can be specified on the command-line which command must be used
to start its run-time environment's daemons (called {\tt orted}). In order to
execute these daemons in the cgroup {\tt xp} on all the mounted controllers, the
{\tt mpiexec} command can receive the following option:

\begin{Verbatim}[frame=single,fontsize=\relsize{-1}]
--mca orte_launch_agent 'cgexec -g *:xp orted' 
\end{Verbatim}

Otherwise, if we want to run the application processes in the aforementioned
control group, the parallel program can be executed by the {\tt cgexec} command:

\begin{Verbatim}[frame=single,fontsize=\relsize{-1}]
mpiexec  -n 5 cgexec -g *:xp mpiprogram
\end{Verbatim}

\section{Implementing failure scenarios with SystemTap}
\label{sec:scenarios}

In this section, we present how failure scenarios can be implemented
using SystemTap. We give two examples: a deterministic scenario
(section \ref{sec:scenarios:determ}) and a probabilistic one (section
\ref{sec:scenarios:probab}). 

A process can be killed at kernel-level by canceling it (and freeing
it) from the scheduler. When the kernel is about to schedule a
process, it calls the {\tt context\_switch} function. As explained in
section \ref{sec:fist:stap}, a \emph{Jprobe} can be inserted when this
function is entered. Then, we can see which control group the process
belongs to (see section \ref{sec:fist:cgroups} and, based on the
failure scenario, decide to kill or not the process. 

In figure \ref{fig:stap:findcgroup} we give an example of a C function
that can be compiled by SystemTap to find out which cgroup a process
belongs to. Based on the unique {\tt pid} of the process, the {\tt
  task\_cgroup} function gets the cgroup of this given task.

\begin{figure*}[ht!]
  \begin{center}
    \begin{minipage}{\linewidth}
      \footnotesize
      \begin{framed}
\begin{verbatim}
function find_cgroup:string(task:long) %{
  struct cgroup *cgrp;
  struct task_struct *tsk = (struct task_struct *)((long)THIS->task);

  /* Initialize with an empty value */
  strcpy(THIS->__retvalue, "NULL");

  cgrp = task_cgroup(tsk, cpu_cgroup_subsys_id);
  if (cgrp)
    cgroup_path(cgrp, THIS->__retvalue, MAXSTRINGLEN);
%}
\end{verbatim}
      \end{framed}
      \normalsize
    \end{minipage}
    \caption{\label{fig:stap:findcgroup}Finding out which cgroup a process
      belongs to.}
  \end{center}
\end{figure*}

SystemTap modules can call functions defined in the kernel. Hence, the
{\tt free\_task} function can be called to cancel and free a process
at the moment when it is about to be scheduled. We give an example of
a piece of code that kills a process calling this function from a
SystemTap module in figure \ref{fig:stap:assassinat}. Therefore, the
process is canceled by the scheduler, but the operating system cannot
do anything else. The I/O structures (\eg network sockets) remain
open, like with any real-life failure. 

\begin{figure*}[h!t]
  \begin{center}
    \begin{minipage}{\linewidth}
      \footnotesize
      \begin{framed}
\begin{verbatim}
function skip_task(task:long) %{
  struct task_struct *tsk = (struct task_struct *)((long)THIS->task);
  free_task( tsk );
%}
\end{verbatim}
      \end{framed}
      \normalsize
    \end{minipage}
    \caption{\label{fig:stap:assassinat}Killing a process by freeing the
      corresponding task in the kernel's scheduler.}
  \end{center}
\end{figure*}

\subsection{Deterministic failure scenarios}
\label{sec:scenarios:determ}

We can imagine the following deterministic failure scenario: after a
certain time {\tt TIMEOUT}, the process is killed. Hence, whenever the
process is about to be scheduled, the SystemTap module needs to
determine for how long it has been running. This information can be
obtained from a field of the {\tt task\_struct} data structure used by
the kernel (see section \ref{sec:fist:cgroups} for more information). 
On recent versions of the Linux kernel, a tapset function provides
this information. 

Figure \ref{fig:scenarios:determ} gives an excerpt of what can be used
by SystemTap. The two functions {\tt task\_stime\_} and {\tt
  task\_utime\_} return respectively the system time and the user time
spent by the process, obtained from the internal {\tt task\_struct}
data structure. When the {\tt context\_switch} function is called, the
module finds out which cgroup the process to be scheduled belongs
to. If it belongs to the {\tt xp} cgroup, the process is concerned by
failure injection. The module finds out for how long the process
has been running. If it has been running for longer than {\tt
  TIMEOUT}, the process is killed.

\begin{figure*}
  \begin{center}
    \begin{minipage}{\linewidth}
      \footnotesize
      \begin{framed}
\begin{verbatim}
function task_stime_:long(task:long){
  if (task != 0)
    return @cast(task, "task_struct", "kernel<linux/sched.h>")->stime;
  else
    return 0;
}

function task_utime_:long(task:long){
  if (task != 0)
    return @cast(task, "task_struct", "kernel<linux/sched.h>")->utime;
  else
    return 0;
}

probe kernel.function("context_switch") {
  cgroup = find_cgroup($next)
  if ( cgroup == "/xp" ){
    t = task_stime_( $next ) + task_utime_( $next )
    if( t > TIMEOUT ) {
      skip_task( $next )
    }
  }
}
\end{verbatim}
      \end{framed}
      \normalsize
    \end{minipage}
    \caption{\label{fig:scenarios:determ}Deterministic failure injection
      scenario: after a certain time TIMEOUT, the process is killed.}
  \end{center}
\end{figure*}

\subsection{Probabilistic failure scenarios}
\label{sec:scenarios:probab}

We can imagine the following probabilistic failure scenario: every time
a process is scheduled, it has one chance out of two (likelihood of
50\%) to be killed. Hence, whenever the process is about to be
scheduled, the SystemTap module generates a random number in $[0,2[$,
    and if this number is larger or equal to 1, then the process is
    killed. 

Figure \ref{fig:scenarios:probab} gives an excerpt of what can be used
by SystemTap. When the {\tt context\_switch} function is called, the
module finds out which cgroup the process to be scheduled belongs
to. If it belongs to the {\tt xp} cgroup, SystemTap generates a random
integer in $[0,2[$ using the function {\tt randint}. If this integer
    is larger or equal to 1, the process is killed.

\begin{figure}
  \begin{center}
    \begin{minipage}{.9\linewidth}
      \footnotesize
      \begin{framed}
\begin{verbatim}
probe kernel.function("context_switch") {
    cgroup = find_cgroup($next)
    if ( cgroup == "/xp" ){
        r = randint( 2 )
        if( r >= 1 ) {
            skip_task( $next )
        }
    }
}
\end{verbatim}
      \end{framed}
      \normalsize
    \end{minipage}
    \caption{\label{fig:scenarios:probab}Probabilistic failure injection
      scenario: every time it is about to be scheduled, the process has a
      certain probability to be killed.} 
  \end{center}
\end{figure}
